\documentclass[sigconf]{acmart}
\pdfoutput=1
\AtBeginDocument{%
  \providecommand\BibTeX{{%
    \normalfont B\kern-0.5em{\scshape i\kern-0.25em b}\kern-0.8em\TeX}}}
\usepackage{ifpdf} \ifpdf \pdfoutput=1 \else \pdfoutput=0 \fi
\copyrightyear{2019} 
\acmYear{2019} 
\setcopyright{acmcopyright}
\acmConference[e-Energy '19]{Proceedings of the Tenth ACM International Conference on Future Energy Systems}{June 25--28, 2019}{Phoenix, AZ, USA}
\acmBooktitle{Proceedings of the Tenth ACM International Conference on Future Energy Systems (e-Energy '19), June 25--28, 2019, Phoenix, AZ, USA}
\acmPrice{15.00}
\acmDOI{10.1145/3307772.3328301}
\acmISBN{978-1-4503-6671-7/19/06}
\usepackage{algorithm}
\usepackage[noend]{algpseudocode}

\usepackage{algorithmwh}
\usepackage{optidef}
\usepackage{graphicx}
\usepackage{caption}
\usepackage{subcaption}
\usepackage{tabularx}
\usepackage[flushleft]{threeparttable}
\begin{document}

%
\title{Structured Dictionary Learning for Energy Disaggregation }

%
\author{Shalini Pandey}

\email{pande103@umn.edu}
\affiliation{%
  \institution{University of Minnesota}
  \city{Twin Cities}
  \state{MN}
  \postcode{55414}
}

\author{George Karypis}

\email{karypis@umn.edu}

\affiliation{%
  \institution{University of Minnesota}
  \city{Twin Cities}
  \state{MN}
  \postcode{55414}
}

\begin{abstract}

The increased awareness regarding the impact of energy consumption on the environment has led to an increased focus on reducing energy consumption. Feedback on the appliance level energy consumption can help in reducing the energy demands of the consumers. Energy disaggregation techniques are used to obtain the appliance level energy consumption from the aggregated energy consumption of a house. These techniques extract the  energy  consumption  of  an  individual  appliance  as  features and hence face the challenge of distinguishing two similar energy consuming devices. To address this challenge we  develop methods that leverage the fact that some devices tend to operate concurrently at specific operation modes. The aggregated energy consumption patterns of a subgroup of devices allows us to identify the concurrent operating modes of devices in the subgroup. Thus, we design hierarchical methods to replace the task of overall energy disaggregation among the devices with a recursive disaggregation task involving device subgroups. Experiments on two real-world datasets show that our methods lead to improved performance as compared to baseline. One of our approaches, Greedy based Device Decomposition Method (GDDM) achieved up to $23.8\%$, $10\%$ and $59.3\%$ improvement in terms of micro-averaged $f$ score, macro-averaged $f$ score and Normalized Disaggregation Error (NDE), respectively. 
\end{abstract}

\maketitle

\section{Introduction}
The residential sector consumes about one-third of the total electricity in the United States and thus, significant opportunities exist to reduce these energy demands~\cite{dietz2009household}. 
Consumers are often unaware as to which appliances consume most of the energy ~\cite{gardner2008short, attari2010public}, and which actions have the greatest savings potential. Individual appliance level energy consumption provides feedback to consumers to improve their energy consumption behavior, detect malfunctioning devices and forecast demand \cite{froehlich2011disaggregated}. But currently, there is no system which can inform consumers about their appliance level energy consumption. As an example, consider consumers receiving a shopping bill with a single figure and being asked to spend less on the next shopping trip. It would be extremely difficult, if not impossible, for consumers to adjust their spending habits without knowing the cost of individual items. Advanced Metering Technique~\cite{szydlowski1993advanced} records the total energy consumption of houses in real time. This gives us an opportunity to perform energy disaggregation which is the task of decomposing the entire energy consumption of a house into appliance level energy consumption.   \par 

Previous approaches that attempt to solve the energy disaggregation problem include deep neural networks~\cite{kelly2015neural}, $k$-nearest neighbours~\cite{batra2016gemello}, discriminative sparse coding~\cite{kolter2010energy}, multilabel classification~\cite{tabatabaei2017toward},  and matrix factorization~\cite{batra2017matrix}. These approaches extract the energy consumption of each appliance as features and use them for decomposing the aggregated energy consumption. However, they face the challenge of correctly disaggregating energy in scenarios where the energy consumption of different devices is similar. Some methods~\cite{elhamifar2015energy,zhong2014signal,tomkins} try to address this challenge by extracting handcrafted features such as duration of usage of an appliance, prior knowledge of appliance concurrence, and average energy consumption in different time intervals. But these methods incorporate extra constraints to take into account the above mentioned features which causes an increased complexity of solving the problem. 
Deep Sparse Coding based Recursive Disaggregation Model (DSCRDM)~\cite{dong2013deep} attempts to solve this problem by capturing the devices which are co-used. 
 However, its dictionary elements do not necessarily represent the energy consumption pattern of devices at specific operation modes, leading to degraded disaggregation performance ~\cite{elhamifar2015energy}. Also, DSCRDM is computationally expensive as it solves non-convex optimization problems at each recursive step. \par

 In our work, we present an extension to the Powerlet based energy disaggregation (PED)~\cite{elhamifar2015energy} to address the limitations mentioned above. PED is a dictionary learning method which captures the different power consumption patterns of each appliance as representatives (used as dictionary atoms) and then estimates a combination of these representatives that best approximates the observed aggregated power consumption. However, it faces the limitation of determining which appliance is operating when devices have similar representatives. For handling such cases, we leverage the fact that some devices instead of merely co-occurring together, operate at specific operation modes (the various modes in which a device can operate) for performing a certain task. To distinguish between two similar power consuming devices, we develop methods that can extract the concurrent operation of different devices without using any extra information. 
\par
 Our approaches extract the representatives among the aggregated power consumption of the device set. If devices in the set co-occur at a specific operation mode then the representative power consumption of that operation mode is one among the extracted representatives. The device set is decomposed level-wise into two partitions each containing equal number of devices (or one partition containing one more device than the other if the number of devices is odd) recursively until only one device is left. Organizing the devices in this fashion creates a binary tree, which induces a hierarchical clustering of the devices. The leaves of that tree correspond to the individual devices and the interior nodes correspond to subsets of devices whose cardinality increases as
we move up the tree. Instead of performing the disaggregation among all the devices together, we decompose the task as a recursive one where the power consumption of subsets of devices is estimated at every level from top to bottom. 
  \par
 We investigated two different heuristics for building the hierarchical tree: Greedy based Device Decomposition Method (GDDM) and Dynamic Programming based Device Decomposition Method (DPDDM). 
To perform disaggregation, we formulate our problem as Semi Definite Programming (SDP) using ~\cite{park2017semidefinite} and solve it using Alternate Direction Method of Multipliers (ADMM)~\cite{boyd2011distributed}, which is more scalable than the optimization approach used in PED. Experiments on two real-world datasets show that our methods lead to improved performance as compared to baseline. GDDM achieved up to $23.8\%$, $10\%$ and $59.3\%$ improvement in terms of micro-averaged $f$ score, macro-averaged $f$ score and Normalized Disaggregation Error (NDE), respectively. \par
\begin{table*}
\caption{Notation.}

\begin{tabular}{ll}
\toprule
Notation & Description \\
\hline
$T$ & Total timestamp for which aggregated energy was recorded on the meter\\
$L$ & The number of devices in the house \\
$x_i(t)$ & The energy consumption of device $i$ at timestamp $t$\\
$\bar{x}(t)$ & The aggregated energy consumption at timestamp $t$\\
$B$ & The dictionary learned for performing disaggregation \\
$B_i$ & The dictionary of $i$th device\\
$B_{ij}$ & $j$th powerlet of $i$th device\\
\bottomrule
\end{tabular}
\end{table*}


\section{Definitions and Notations}

 In this paper, all vectors are represented by bold lower case letters (e.g., $\boldsymbol{y}$), all matrices are represented by bold upper case letters (e.g., $\boldsymbol{B}$), and sets using upper case caligraphic letters (e.g., $\mathcal{S}$).\par
 \subsection{Problem Definition}
Let $L$ be the number of devices in a house, $x_i(t)$ be the energy consumption of device $i$ at timestamp $t \in \{1, \ldots, T\}$, and $\bar{x}(t) = \sum_{i=1} ^L x_i(t)$ be the aggregated energy consumption of the house at timestamp $t$. 
The goal of energy disaggregation is, given the total energy consumed by a house, $\bar{x} (t)$, at time $t$, identify the energy consumption of every device in the house at that time, i.e., estimate $x_i(t)$ for $i \in \{1, \ldots, L\}$.

\section{Related Work}

Energy disaggregation or non-intrusive load monitoring was initially studied and explored in~\cite{hart1992nonintrusive}. Various data mining and pattern recognition methods have been applied to find the contribution of each appliance in the total energy consumption. Methods such as deep neural networks~\cite{kelly2015neural}, $k$-nearest neighbours~\cite{batra2016gemello}, discriminative sparse coding~\cite{kolter2010energy}, multilabel classification~\cite{tabatabaei2017toward} and matrix factorization~\cite{batra2017matrix} have been used to solve energy disaggregation problem. These methods extract electrical events as features and use them for decomposing the aggregated energy consumption. On the other hand, methods like~\cite{kolter2012approximate,kim2011unsupervised,tomkins,shaloudegi2016sdp} exploit the sequential nature of the aggregated energy consumption data by modeling every appliance as an independent Hidden Markov Model. However, these methods do not focus on learning a structure which decomposes the device set systematically and helps in finding a disaggregation sequence. The disaggregation sequence improves the overall accuracy of the method as is described in ~\cite{dong2013deep}. Another work~\cite{rahimpour2017non}  models the disaggregation problem as a non-negative matrix factorization problem with constraints to reduce the effect of correlation
between atoms in the dictionary. The difference between this method and what we present in our paper is that the grouping effect is induced within the atoms of a particular device only, instead our method induces a grouping effect of different devices in the house.   Another competing approach, seq2point~\cite{zhang2018sequence}, uses a convolution neural network to map between the mains power readings and single appliance power readings. Their networks extract energy consumption values and duration of energy consumption as features but again fail to consider the co-occurrent power consumption of subset of devices.\par
We now describe the two methods that this paper is built upon: 

  \begin{figure}[t!]
  \centering
  \includegraphics[keepaspectratio, width=0.5\textwidth]{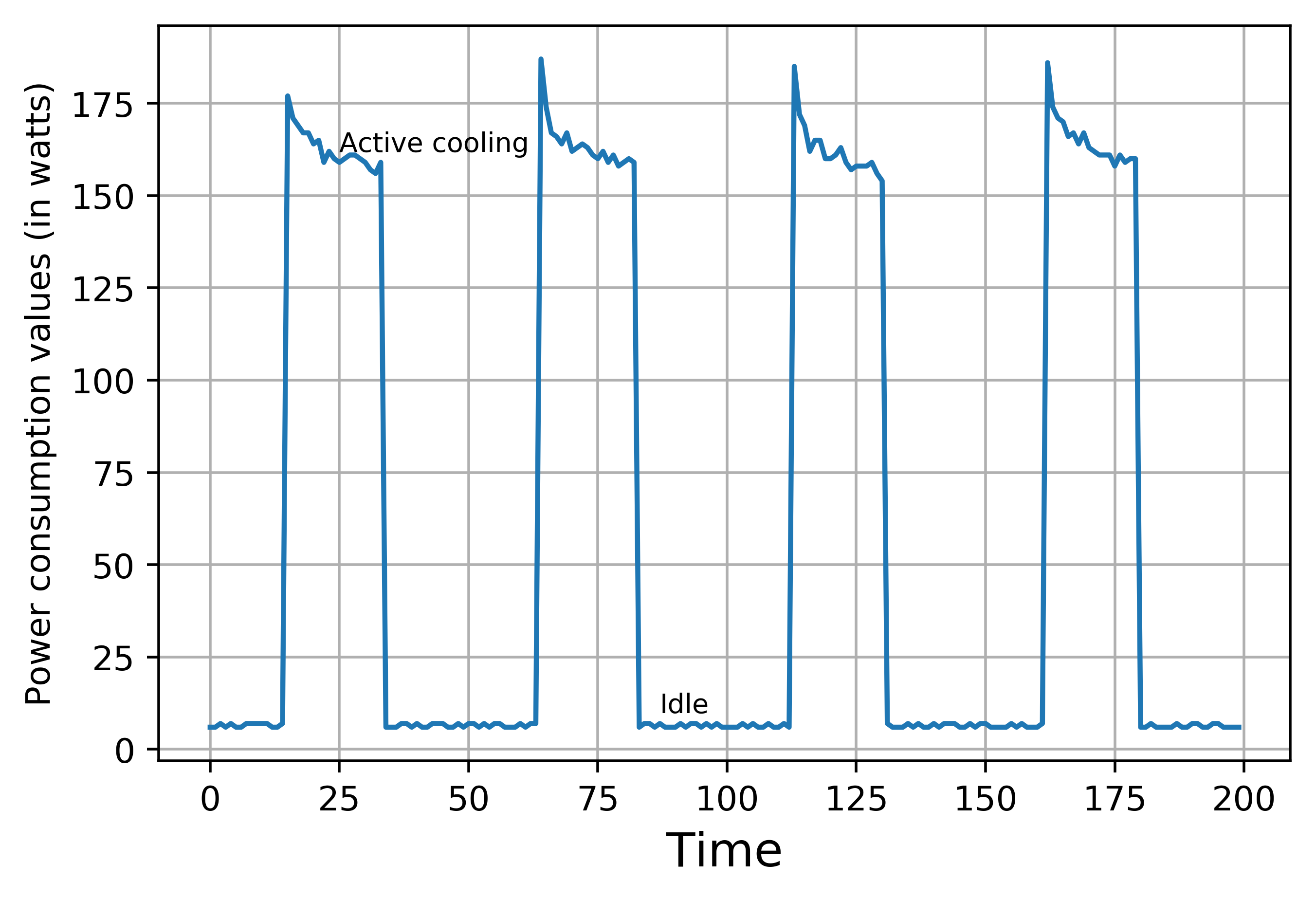}
  \caption{Energy consumption of a refrigerator. Horizontal and vertical axes correspond to time (seconds) and energy consumption (Watts), respectively. The energy consumption at active cooling and idle operation modes are shown. This figure shows that there exists pattern in energy consumption of appliances. }
  \label{fig:energyconsumed}
\end{figure}
\subsection{Powerlet Based Energy Disaggregation (PED)}
PED introduced the concept of powerlets for energy disaggregation. It solves the problem by breaking the task into two phases, extracting the energy consumption pattern of each device and performing the actual disaggregation task among the extracted energy consumption patterns. \par
 The operation modes of a device refer to the various modes in which it can operate based on the energy that it consumes. For example, a microwave can operate in operation modes such as defrost, heat with high power or low power or switched off state. 
 The training phase of PED assumes that each device's operation mode can be mapped to a vector of length $w$, referred to as \emph{\textbf{powerlet}}, which represents
the energy consumption pattern in that operation mode. \par 
 To find the powerlets of a device $i$, the energy consumption of the device in time interval $[t, t+w-1]$ is represented by a $w$-dimensional vector, $\boldsymbol{y}_i(t)$,
where $t \in \{1,2, \ldots, T-w+1\}$. 
 Powerlets for device $i$ are selected from the set of vectors $\mathcal{V} = \{\boldsymbol{y}_i(t), \forall t\in\{1, 2,\ldots T-w+1\}\}$, using a variant of the DS3 algorithm for sequential data \cite{elhamifar2016dissimilarity}. 
Let subdictionary $\boldsymbol{B}_i\in R^{w \times N_i}$ consists of the powerlets of device $i$, represented by a
matrix whose columns are the powerlets and $N_i$ is the number of powerlets of device $i$. The dictionary, $\boldsymbol{B}
\in R^{w\times N}$ where $N = \sum_{i=1}^{L} N_i$ is the set of subdictionaries
of all the devices in the house and is obtained by concatenating the subdictionaries of all the devices, i.e.,

\begin{equation}
    \boldsymbol{B} = [\boldsymbol{B}_1, \ldots, \boldsymbol{B}_L].
\end{equation}

The vector $\boldsymbol{\bar{y}}(t)$ represents the aggregated energy consumption
in time interval $[t, t + w-1$] i.e.,

\begin{equation}
    \boldsymbol{\bar{y}}(t) = [\bar{x}(t), \bar{x}(t + 1), \ldots, \bar{x}(t + w-1) ].
\end{equation}

In its second phase, PED disaggregates the energy by solving the following optimization problem:

  \begin{mini} 
      {\boldsymbol{c}(t)} { \lambda \rho({\boldsymbol{c}(t)}_{t=1}^T)+ \sum_{t=1}^T|| {\boldsymbol{\bar{y}}(t)}-\boldsymbol{B}\boldsymbol{c}(t)||_1 }{}{} 
  \addConstraint{\boldsymbol{c}_i(t)=\{0,1\}^{N_i}}
  \addConstraint{1^T\boldsymbol{c}_i(t) \leq 1}
 \end{mini}

\noindent where $\boldsymbol{c}(t) = [\boldsymbol{c}_1(t), \boldsymbol{c}_2(t), \ldots , \boldsymbol{c}_L(t)] \in\mathbb{R}^N$, $\boldsymbol{c}_i(t)\in\mathbb{R}^{N_i}$ is the coefficient of the powerlets of device $i$, $\lambda$ is a regularization parameter, and $\rho({\boldsymbol{c}(t)}_{t=1}^T)$ indicates the penalty
 associated with the priors such as device-sparsity, knowledge about devices that do or do not work together, and temporal consistency of the disaggregation. 
 The first constraint enforces the selection or not of a powerlet and the second constraint is used to ensure that at most one powerlet is selected for every device. \par
 However, PED does not take into account the concurrent operation modes of devices within the time window $w$. The co-occurrence prior in PED only considers the devices which are used together or which are not used together. 
 Instead, certain devices tend to be used concurrently at specific operation modes, which arises when certain tasks are performed.
\subsection{Recursive dictionary learning for energy disaggregation}
 Deep Sparse Coding based Recursive Disaggregation Model\\ (DSCRDM)~\cite{dong2013deep} exploits a tree structure for the problem of energy disaggregation, such that, at a particular level, one of the nodes contains one device and the other contains the rest. That device is separated from the remainder which minimizes the disaggregation error on the training set.  DSCRDM is based on Discriminative Disaggregation Sparse Coding~\cite{kolter2010energy} (DDSC) to find the dictionary of each appliance. DDSC models the entire energy consumption of each device
 as a sparse linear
combination of the atoms of an unknown dictionary.\par

\begin{figure*}
\centering
\begin{subfigure}[b]{.4\linewidth}
\includegraphics[width=\linewidth]{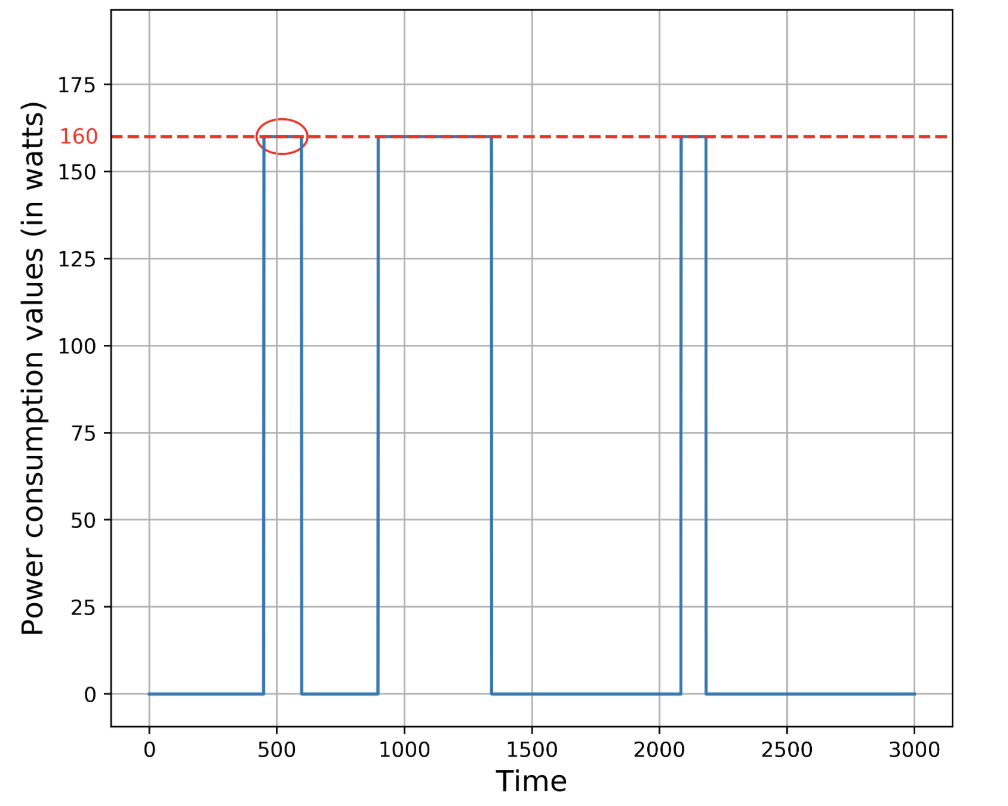}
\subcaption{Energy consumption of stove}\label{fig:mouse}
\end{subfigure}%
\begin{subfigure}[b]{.4\linewidth}
\includegraphics[width=\linewidth]{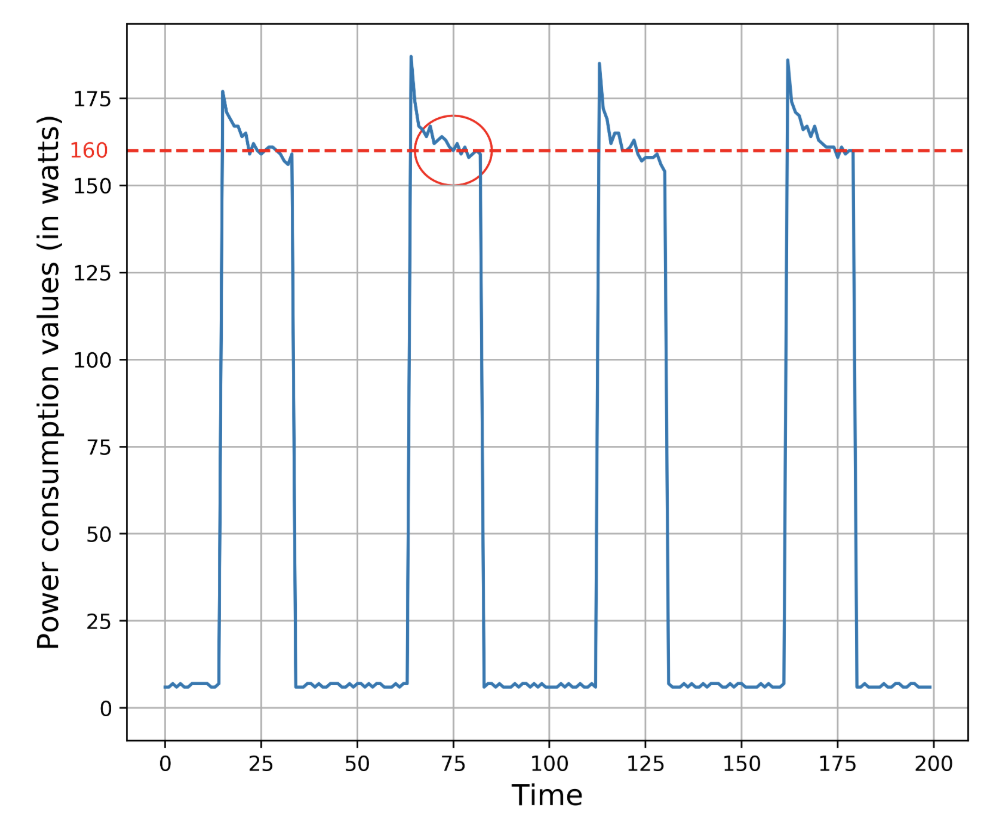}
\subcaption{Energy consumption of refrigerator}\label{fig:gull}
\end{subfigure}
\begin{subfigure}[b]{.4\linewidth}
\includegraphics[width=\linewidth]{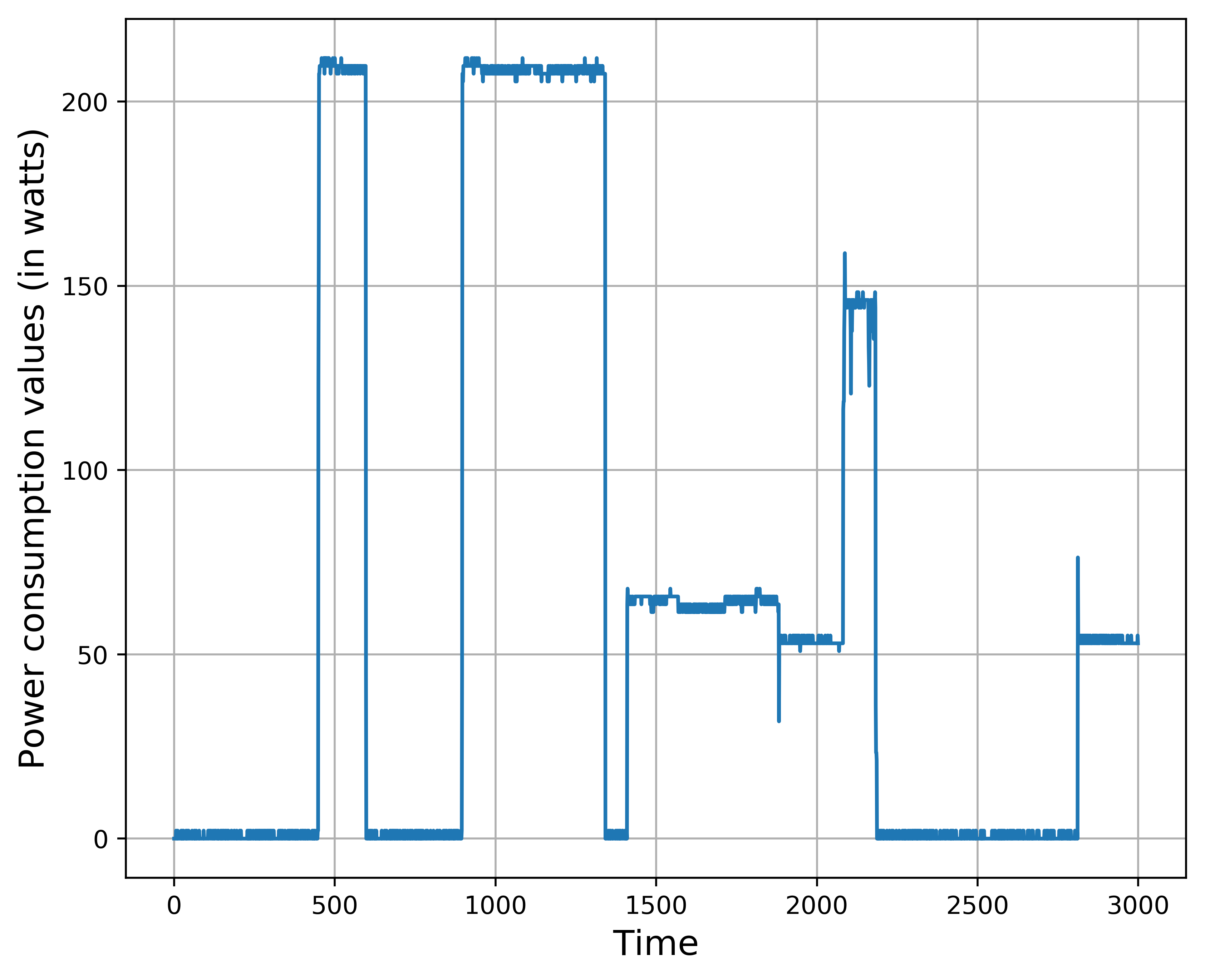}
\centering
\subcaption{Energy consumption of air exhaust}\label{fig:tiger}
\end{subfigure}

\caption{Illustration of devices having a similar energy consumption value.  Both stove and refrigerator show a pattern where
they tend to operate at 160W. The prior approaches may fail to distinguish between the power consumption of the
two devices. We can resolve this by using the fact that the air exhaust is on whenever the stove is in use.}
\label{devices}
\end{figure*}

However, the dictionary atoms used in DDSC are arbitrary vectors which do not correspond to the energy consumption pattern of specific operation modes of the device. This degrades the disaggregation performance~\cite{elhamifar2015energy}.  Moreover, DSCRDM solves a non-convex problem to find the best split at each node which increases the computational complexity of training the decomposition structure. 
 \par

\section{Proposed Method}
Similar to PED, we follow a two-step procedure for energy disaggregation. Firstly, we learn powerlets of different subsets of devices and the decomposition structure using the training dataset.
Then we decode the aggregated energy consumption among the learned
dictionary atoms at each level of the decomposition structure following an optimization scheme similar to PED. 
In this section, we first provide the motivation for our method followed by the framework we developed for energy disaggregation.

\subsection{Motivation}
The prior approaches for energy disaggregation face the common challenge of disaggregating energy of similar power consuming devices. For example, in Figure ~\ref{devices}
 both stove and refrigerator consume similar power. However, we observe that the air exhaust is usually on when the stove is used. Thus, for performing disaggregation we can consider the aggregated power consumption of air exhaust and stove as they have different power consumption pattern compared to the refrigerator alone. Grouping devices into sets and then performing disaggregation can help in distinguishing two similar power consuming devices. Motivated by this, we develop \textbf{\textit{device decomposition methods}} to find the sequence in which energy is to be disaggregated.   \par
 Another fact to consider is that devices do not just merely co-occur but they co-occur at specific operation modes for performing one task and for another task they can co-occur at different modes. For example, when heating food, a person will start by opening the refrigerator to get the food item, which can potentially transition the fridge to an active cooling operation mode. Then the person puts the item in the microwave and switches it on. Thus, the transition of operation modes of the two devices tends to occur simultaneously. After switching the microwave on, the person puts the left out food back in the fridge which again changes the operation mode of the fridge. However, when a dish is being cooked, some other appliance in the kitchen can be used along with the microwave and fridge. Thus, a set of devices are used at specific modes, for certain tasks (food heating), whereas for other tasks (cooking), another set of devices are used simultaneously. \par

To summarize, it is important to group devices into sets and then perform disaggregation. Additionally, the co-occurence of devices is not specific to two devices merely operating together but the operation modes at which they co-occur is also important. 
\subsection{Learning powerlets}
A powerlet is a vector of length $w$, which is used as a representative for a specific operation mode of a device. In order to find the powerlets, we select representatives among the set of vectors $\mathcal{V} = \{\boldsymbol{y}_i(t), \forall t\in\{1, 2,\ldots T-w+1\}\}$, where $\boldsymbol{y}_i(t)$ is the energy consumption of the device during the time interval $[t, t+w-1]$. The DS3 algorithm employed in PED has high storage and computational complexity~\cite{mavrokefalidis2016supervised}. A simpler clustering based method~\cite{tosic2011dictionary} can be used for selecting the representatives among the set of vectors. Specifically, we use $k$-medoid algorithm because both $k$-medoid and DS3 select representatives from the data to be clustered. 
\subsection{Learning dictionary and device decomposition structure} 
\begin{figure}

\small
\centering

\includegraphics[keepaspectratio, width=0.4\textwidth]{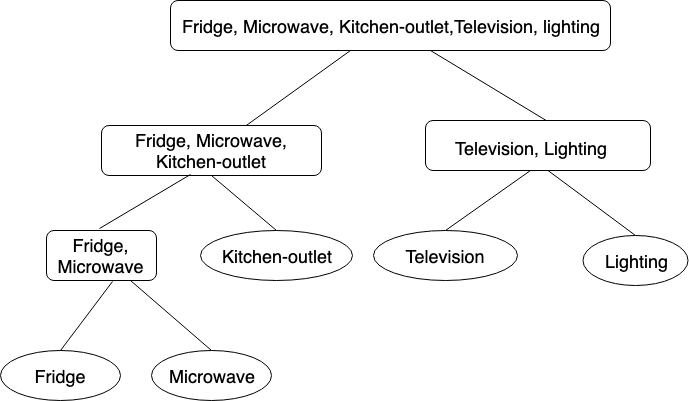}
\caption{An example of binary tree structure of a particular house.}
\label{fig:binary structure}
\end{figure}

We create the device decomposition structure by decomposing the device set recursively into two equal halves (or one half with one device more than the other if the number of devices is odd) until only one device remains. This creates a binary tree structure containing sets of devices at every node. We choose to use binary tree as device decomposition structure because of its simplicity. For understanding purpose, we define a pseudo device as a hypothetical device whose energy consumption is the same as the aggregated energy consumption of the devices in the set, i.e., energy consumption of a pseudo device at time $t$, $x_{\rho}(t)$ is calculated as, 
\begin{equation}
  x_{\rho}(t) = \sum_{i \in \mathcal{S}}x_i(t).
\end{equation}
We extract the powerlets of the pseudo devices at each node. If devices in the set co-occur at a specific operation mode then the representative power consumption of that operation mode is one among the extracted representatives.  
We perform energy decomposition in a recursive manner, starting from the root node and disaggregating the aggregated energy assigned at every node between its two children. Figure~\ref{fig:binary structure} shows an example of the device decomposition structure.\par
In order to maximize the accuracy of decomposing the energy signal, we focus on making the powerlets of devices used for disaggregation as dissimilar as possible. A key parameter for constructing the device decomposition structure is the method that is used to partition the set of devices at each node of the tree.


  
\subsubsection{Greedy based Device Decomposition Method (GDDM):}
This approach splits the set of devices into two groups by maximizing the dissimilarity between powerlets at each level. 
\par
Let $S_1$ and $S_2$ be the two subsets of devices created from the set of devices. Let $B_1$ and $B_2$ be the powerlets for these two  pseudo devices. Dissimilarity of $S_1$ and $S_2$ is the distance between the pair of closest powerlets, one from each of the pseudo device, which is defined as,  
\begin{equation}
     dissimilarity(S_1, S_2) = \min_{p,q} ||\boldsymbol{B}_{1 p}-\boldsymbol{B}_{2q}||_1,
\end{equation}
Since the disaggregation is performed using the powerlets of the pseudo devices, we want the resultant powerlets to be as dissimilar as possible. Hence, we maximize the dissimilarity between the two pseudo devices. We explore two different ways of splitting the device set greedily. \\
\textbf{Equi-sized partition:} To find equi-sized partitions, we start by randomly dividing the device set equally among the two nodes (or with one node having one more device than the other if the number of devices in the node is odd). 
We find a device, $d_1$ from the first node, which when moved to the second node leads to the maximum increase in the dissimilarity between the resultant pseudo devices. 
Similarly we find a device, $d_2$, from the resultant second node which when moved to the first node increases the dissimilarity between the resultant pseudo devices. We then move $d_1$ to the second node and $d_2$ to the first. We repeat this step until no such device is found which when moved increases the dissimilarity between the pseudo devices. Algorithm 1 describes how equi-sized partition is performed using GDDM. \\
\begin{algorithm}[H]
\begin{code}
{\bf Input:} $S:$ The set of devices to be partitioned \\
\uln \textbf{begin}\\
\uln \quad$l\leftarrow $number of devices in $\mathcal{S}$\\ 
\uln \quad Iterate until $d_1$ or $d_2$ is NULL: \\
\uln \quad\quad$\mathcal{S}_1 \leftarrow \text{$l$/2 randomly selected devices} \in \mathcal{S}.$\\
\uln \quad\quad$\mathcal{S}_2 \leftarrow \mathcal{S}\setminus \mathcal{S}_1$ \\
\uln \quad\quad$d_1 \leftarrow split(\mathcal{S}_1, \mathcal{S}_2)$\\
\uln \quad\quad$d_2 \leftarrow split(\mathcal{S}_2\cup d_1, \mathcal{S}_1 \setminus d_1)$\\
\uln \quad\quad$\mathcal{S}_1 \leftarrow \mathcal{S}_1 \cup d_2$\\
\uln \quad\quad$\mathcal{S}_2 \leftarrow \mathcal{S}_2 \cup d_1$\\
\uln \quad Return $\mathcal{S}_1, \mathcal{S}_2$
\end{code}
\label{alg:main}
\caption{Greedy based Maximizing Inter-set Dissimilarity Method (Equi-sized Partition)}
\end{algorithm}
\begin{algorithm}[H]
\begin{code}
{\bf Procedure:} split($\mathcal{S}_1$, $\mathcal{S}_2$) \\
{\bf Input:} $S:$ The set of devices to be partitioned \\
\uln \textbf{begin}\\
\uln \quad \text{for} $ i \in \mathcal{S}_1$:\\
\uln \quad\quad $\boldsymbol{B}_{\rho}^1 \leftarrow$ Powerlets of pseudo device $\mathcal{S}_1\setminus i$\\
\uln \quad\quad $\boldsymbol{B}_{\rho}^2 \leftarrow$ Powerlets of $\mathcal{S}_2 \cup i$\\
\uln \quad\quad \emph{CM}(i)= $\min_{p,q} ||\boldsymbol{B}_{{\rho}^2 p}-\boldsymbol{B}_{{\rho}^1 q}||_1 $\\
\uln \quad$d \leftarrow \max_i$ \emph{CM}(i) \\
\uln \quad $B_{S_1} \leftarrow$ powerlets of pseduo device $\mathcal{S}_1$ \\
\uln \quad$B_{S_2} \leftarrow$ powerlets of pseduo device $\mathcal{S}_2$\\
 \uln\quad\emph{CM}(d) $>  \min_{p,q}||B_{{S_1} p}-B_{{S_2} q}||$:\\
 \uln \quad\quad Return NULL\\
 \uln \quad else:\\
 \uln \quad\quad Return $d$\\
\end{code}
\label{alg:main}

\end{algorithm}
\textbf{1-vs-rest partition:}
1-vs-rest partitioning refers to the partitioning of a set of devices such that one of the subsets contains one device and the other contains the rest. 
    For 1-vs-rest partitioning, we find a device $d$ such that its powerlets are most dissimilar to the powerlets of the pseudo device created by the remainder devices in the set. We then create two subsets - one containing the device $d$ and the other containing the rest. This step is repeated recursively until only one device is the left in the subset. Algorithm 2 describes 1-vs-rest partitioning using greedy mechanism. \\
    \begin{algorithm}[H]
\begin{code}
{\bf Input:} $S:$ The set of devices to be partitioned \\
\uln \textbf{begin}\\
\uln \quad \text{for} $ d \subset  \mathcal{S} $ \\
\uln \quad\quad $\boldsymbol{B}_{d} \leftarrow$ Powerlets of device $d$\\
\uln \quad\quad $\boldsymbol{B}_{\rho} \leftarrow$ Powerlets of $\mathcal{S}\setminus d$\\

\uln \quad\quad $CM(d)=   \min_{p,q} ||\boldsymbol{B}_{{\rho} p}-\boldsymbol{B}_{{d} q}||_1 $\\

\uln \quad Return $\text{device}=\argmax_i CM(i)$
\end{code}
\caption{Greedy based Device Decomposition Method (1-vs-rest)}
\label{alg:main}
\end{algorithm}    
\subsubsection{Dynamic Programming based Device Decomposition Method (DPDDM):}
 In case of GDDM, we split a set of devices such that the dissimilarity between the powerlets of the pseudo devices corresponding to the children node can be maximized. Sometimes, splitting of devices to maximize the dissimilarity of only the resulting pseudo devices can result in children nodes containing devices with similar powerlets. For example, consider a house with devices $d_1$, $d_2$, and $d_3$, such that $d_1$ and $d_2$ have similar power consumption. Greedy based methods can result in a decomposition structure that it predicts the power consumption of $d_3$ followed by $d_1$ and $d_2$. Since it is difficult to distinguish the consumption pattern of $d_1$ and $d_2$, disaggregating energy between them can lead to error. However, DPDDM tries to find the optimal decomposition structure to address this limitation. The split is made by taking into consideration the overall dissimilarity between the powerlets that construct dictionary after any split. In DPDDM, we maximize the following objective function to find the decomposition structure,
\begin{equation}
    \sum_{\rho=1}^{N} |{\mathcal{S_\rho}}|^{\alpha} \min_{p,q} ||\boldsymbol{B}^1_{\rho p}-\boldsymbol{B}^2_{\rho q}||_1,
\end{equation}
where $N$ is the total number of non-terminal nodes in the binary tree structure, $\mathcal{S}_\rho$ is the device set corresponding to the node $\rho$, $\boldsymbol{B}^1_\rho$ and $\boldsymbol{B}^2_\rho$ correspond to the subdictionaries of the pseudo devices of the left and right child of the node $\rho$, respectively, and $\alpha$ is a tunable parameter. Note that Equation (6) is a weighted combination of dissimilarity of powerlets used for disaggregation at all the levels in the decomposition structure. Since an error made in the top nodes of the decomposition structure propagates down the tree, we assign more weight to dissimilarity between powerlets of pseudo devices occurring in the top nodes than the bottom ones. The tree that gives the maximum value of the objective function is selected as the decomposition structure. To optimize the above objective function, we use the following recurrence relation while splitting the nodes, 
\begin{equation}
    C_{\mathcal{S}} = \max_{{\mathcal{P}}, |{\mathcal{P}}|={{|{\mathcal{S}}|}/2}} C_{\mathcal{P}}+C_{\mathcal{S}/\mathcal{P}}+ |{\mathcal{S}}|^{\alpha} \min_{{p},{q}} ||\boldsymbol{B}^1_{\rho p}-\boldsymbol{B}^2_{\rho q}||_1,
\end{equation}
where $C_{\mathcal{S}}$ is the optimum cost of the tree rooted at the node corresponding to device set $\mathcal{S}$, $B_{\rho}^1$ is the subdictionary of the pseudo device of $\mathcal{P}$ and $B_{\rho}^2$ is the subdictionary of the pseudo device of $\mathcal{S}/\mathcal{P}$. 
Algorithm 3 describes Dynamic Programming based Maximizing Inter-set Dissimilarity method.

\begin{algorithm}[H]
\begin{code}
{\bf Input:} $S:$ The set of devices to be partitioned \\
\uln \textbf{begin}\\
\uln \quad \text{for} $ \mathcal{P} \subset  \mathcal{S} $ such that $|\mathcal{P}|=|\mathcal{S}|/2$:\\
\uln \quad\quad $\boldsymbol{B}_{\rho}^1 \leftarrow$ Powerlets of pseudo device $\mathcal{P}$\\
\uln \quad\quad $\boldsymbol{B}_{\rho}^2 \leftarrow$ Powerlets of $\mathcal{S}\setminus \mathcal{P}$\\
\uln \quad\quad $C_\mathcal{S}^1= $ DPDDM($\mathcal{P}$)\\
\uln \quad\quad $C_S^2=$  DPDDM($\mathcal{S}\setminus\mathcal{P}$)\\
\uln \quad\quad $C_\mathcal{P}= C_\mathcal{S}^1+C_\mathcal{S}^2$ \\ \quad \quad \quad \quad $+  |\mathcal{S}|^\alpha|\min_{p,q} ||\boldsymbol{B}^2_{{\rho} p}-\boldsymbol{B}^1_{{\rho} q}||_1 $\\

\uln \quad$\mathcal{Q} = \argmax_{\mathcal{P}}C_\mathcal{P}$ \\
\uln \quad Return $\mathcal{Q}$, $\mathcal{S}\setminus \mathcal{Q}$
\end{code}
\caption{Dynamic Programming based Device Decomposition Method (DPDDM)}
\label{alg:main}

\end{algorithm}

\subsection{Decomposing the energy signal}

In order to disaggregate the energy, we start from the root node containing all the devices and approximate the aggregated energy consumption of the devices that belong to the child nodes. This step is performed recursively until we reach the leaf nodes which contain only one device. Thus, we modify the entire task of decomposing the energy signal into a recursive task. Note that, this is not the same as expressing Equation (3) recursively because the dictionaries involved in performing the disaggregation at each recursion is derived from the powerlets of the corresponding pseudo devices.
 Specifically, at every node, $\rho$, of the tree, the dictionary, $\boldsymbol{B}_{\rho} $ to be used for disaggregating the energy consumption is obtained by concatenating the subdictionaries of the pseudo devices of its two children, i.e., $B_\rho = [B^1_\rho, B^2_\rho]$. 
Since a (pseudo) device can be operating in one of the operation modes, we select exactly one powerlet from the dictionary of the powerlets for every (pseudo) device.  We solve the following optimization problem to obtain the energy consumption of the (pseudo) devices at every step: 

\begin{mini}
  {\boldsymbol{c}(t)} { \sum_{t=1}^T||(\boldsymbol{\bar{y}}_\rho(t)-\boldsymbol{B}_\rho\boldsymbol{c}(t))||_2 }{}{}
  \addConstraint{\boldsymbol{c}_i(t)=\{0,1\}^{N_i}} \addConstraint{1^T\boldsymbol{c}_i(t) = 1},
 \end{mini}
 
\noindent where $\boldsymbol{\bar{y}}_{\rho}(t)$ is the aggregated energy consumption of the
devices that belong to node $\rho$. The last two constraints ensure the selection of exactly one powerlet per device\footnote{Note that we have added the powerlets of the operation mode corresponding to switch off state in the dictionary.}. 
\par
Using decomposition structure for performing disaggregation of energy signal provides speedup as the time complexity of solving the non convex problem using standard solvers is exponential in the number of devices, $k$. By decomposing the problem into recursive, independent subproblems with less number of variables at each optimization, we reduce the time complexity to linear with respect to number of devices. \par

To solve Equation (8) we used ADMM based method as it is more computationally efficient~\cite{park2017semidefinite} and scalable compared to integer programming based method used in PED. Note that, even though ADMM is scalabale with time complexity $\mathcal{O}(k^3)$, our decomposition methods reduced the time complexity to $\mathcal{O}(k)$.\par
In order to reframe this problem in a general form, assume $\boldsymbol{c}(t) \in \mathbb{R}^N$ as $\boldsymbol{x}$, where $N$ is the sum of number of the powerlets of the two (pseudo) devices belonging to the child nodes, $\bar{\boldsymbol{y}}_{\rho}(t)$ as $\boldsymbol{b}$ and $\boldsymbol{B}_\rho$ as $\boldsymbol{A}$. All the linear constraints are combined such that matrix $\boldsymbol{P} \in \mathbb{R}^{2\times N}$ has $i$th row element, $e_i\in \mathbb{R}^N$ whose entries for the powerlets of pseudo (device) $i$ are one and the rest are zeros, $\boldsymbol{q}\in \mathbb{R}^2$ is a vector consisting of all ones.

\begin{mini}
  {\boldsymbol{x}} {||\boldsymbol{Ax}-\boldsymbol{b}||_2 }{}{}
  \addConstraint{x_i(x_i-1)=0}
  \addConstraint{\boldsymbol{P}^T\boldsymbol{x} = \boldsymbol{q}}.
 \end{mini}
 
\noindent 
Following the idea of SDP relaxation for integer programming problem~\cite{park2017semidefinite}, we can reformulate the above optimization as\footnote{The only modification is that we keep the equality constraint which is not present in the equation given in ~\cite{park2017semidefinite}.}:

 \begin{mini}
{\boldsymbol{X}, \boldsymbol{x}} {Tr({\boldsymbol{A^TA}}{\boldsymbol{X}})-2*\boldsymbol{b}^T\boldsymbol{A}\boldsymbol{x}}{}{}
\addConstraint{diag(\boldsymbol{X}) \geq \boldsymbol{x}}
 \addConstraint{\Big[  \begin{array}{cc}
1 & \boldsymbol{x}^T \\
\boldsymbol{x} & \boldsymbol{X} \\  
\end{array} \Big] \succeq 0}{}
 \addConstraint{\boldsymbol{P}^T\boldsymbol{x} = \boldsymbol{q}}{}.
  \end{mini}

By introducing \( \hat{\boldsymbol{X}} = \displaystyle \Big[ \begin{array}{cc} 
1 & \boldsymbol{x}^T \\
\boldsymbol{x} & \boldsymbol{X} \\  
 \end{array} \Big] \) we can represent Equation (10) in a more compact form as

 \begin{mini}
 {\hat{\boldsymbol{X}}} {Tr({\hat{\boldsymbol{A}}}{\boldsymbol{\hat{X}}})}{}{}
\addConstraint{\boldsymbol{\hat X} \succeq 0}
\addConstraint{{\mathcal{A}}(\boldsymbol{\hat{X}}) = {{\boldsymbol{\tilde{b}}}}},
  \end{mini}

\noindent where \( \hat{\boldsymbol{A}}=\displaystyle{\Big[  \begin{array}{cc}
0 & -\boldsymbol{b^TA} \\
-\boldsymbol{(b^TA)^T} & \boldsymbol{A^TA} \\  
\end{array} \Big]} \in \mathbb{S}_{+}^{N+1}, \tilde{\boldsymbol{b}} \in \mathbb{R}^m \) and ${\mathcal{A}}: \mathbb{S}_{+}^{N+1} \rightarrow \mathbb{R}^m$ is a linear map~\cite{wen2010alternating},  $m$ is the number of equality constraints.
 In addition, we impose a non-negativity constraint on the matrix $\hat{\boldsymbol{X}}$. The final equation then becomes

 \begin{mini}
 {\hat{\boldsymbol{X}}} {Tr({\boldsymbol{\hat{A}}}{\boldsymbol{\hat{X}}})}{}{}
\addConstraint{\boldsymbol{\hat X} \succeq 0}
\addConstraint{{\mathcal{A}}(\boldsymbol{\hat{X}}) = {\tilde{\boldsymbol{b}}}},
  \end{mini}

\noindent where \( \hat{\boldsymbol{A}}=\displaystyle{\Big[  \begin{array}{cc}
0 & -\boldsymbol{b^TA} \\
-\boldsymbol{(b^TA)^T} & \boldsymbol{A^TA} \\  
\end{array} \Big]} \in \mathbb{S}_{+}^{N+1}, \tilde{\boldsymbol{b}} \in \mathbb{R}^m \) and ${\mathcal{A}}: \mathbb{S}_{+}^{N+1} \rightarrow \mathbb{R}^m$ is a linear map~\cite{wen2010alternating},  $m$ is the number of equality constraints.
 In addition, we impose a non-negativity constraint on the matrix $\hat{\boldsymbol{X}}$. The final equation then becomes

 \begin{mini}
{\hat{\boldsymbol{X}}} {Tr(\hat{\boldsymbol{A}}\hat{\boldsymbol{X}})}{}{}
 \addConstraint{\hat{\boldsymbol{X}} \succeq 0}, {\hat{\boldsymbol{X}} \geq 0}
 \addConstraint{{\mathcal{A}} (\hat{\boldsymbol{X}}) = {\tilde{\boldsymbol{b}}}}.
  \end{mini}

\par To apply ADMM on Equation (11), we use Moreau-Yosida quadratic regularization ~\cite{malick2009regularization}. 

Lagrangian function for Equation (11) can be written as
\begin{multline*}
    \mathcal{L}_\mu=Tr(\hat{\boldsymbol{A}}\hat{\boldsymbol{X}})+ \frac{1}{2\mu}{||\hat{\boldsymbol{X}}-\boldsymbol{S}||}_{F}^{2} +\lambda^T(\tilde{\boldsymbol{b}}-\mathcal{A} (\boldsymbol{X})\\-Tr(\boldsymbol{W}^T\hat{\boldsymbol{X}})-Tr(\boldsymbol{P}^T\hat{\boldsymbol{X}}),
  \end{multline*}
  \noindent where $\lambda \geq 0,\ \boldsymbol{P}\succeq0,\ ,\boldsymbol{W}\geq0$ are dual variables and $\mu > 0$ is a constant.
   
By taking the derivative of ${\mathcal{L}}_\mu$ and computing optimal value of $\hat{\boldsymbol{X}}$, one can derive the standard updates of ADMM solver.\par 
After obtaining optimal, $\boldsymbol{X}^\star$ and $\boldsymbol{x}^\star$ from $\hat{\boldsymbol{X}}$, we use randomized rounding~\cite{park2017semidefinite} for finding an approximate solution, $\boldsymbol{x}^{opt}$ to our original problem   which minimizes the optimization function of Equation (12). We randomly sample $\boldsymbol{z}$ from ${\mathcal{N}}(\boldsymbol{x}^\star,\boldsymbol{X}^\star-\boldsymbol{x}^\star \boldsymbol{x}^{\star T})$. After obtaining $\boldsymbol{z}$, we greedily change its values so that the constraint of selection of exactly one powerlet per device is met. After performing this random sampling for few iterations, we select the optimal $\boldsymbol{z}^\star$ that gives lowest value of objective function as $\boldsymbol{x}^{opt}$.

Since the results at various time instants are independent, we solve each integer quadratic problem sequentially.

\section{Experimental Setup}

\subsection{Datasets}
We evaluate our results on the publicly available Electricity Consumption \& Occupancy (ECO) ~\cite{beckel2014eco} and the UK Domestic Appliance-Level Electricity (UK-DALE) ~\cite{kelly2015uk} datasets.\\
\textbf{ECO}~\cite{beckel2014eco}: This dataset contains the energy consumption of 6 households and the average number of appliances in the households is 7. The data was collected  over a span of 8 months. The meter readings is recorded at a frequency of 1 Hz. We sample the data so that the consecutive energy consumption readings are at an interval of one minute. \\ 
\noindent \textbf{UK-DALE}~\cite{kelly2015uk}:  This dataset monitors $5$ households and the recordings are available every $6$ second. In this dataset, there are varying number of devices from $6$ to $24$. We sample this data also so that the consecutive energy consumption readings are at an interval of one minute.\\ The ground truth data is the actual energy consumption of each appliance. We use the readings with mains label in the dataset as the aggregated power consumption value. \\
  

\subsection{Evaluation Methodology and Performance Assessment Metrics} 

\textbf{Approaches:} We compare our methods with three other existing methods: PED~\cite{elhamifar2015energy}, Seq2point~\cite{zhang2018sequence} and DSCRDM~\cite{dong2013deep}, which are described in section 3. DSCRDM applies its decomposition learning on DDSC which faces the challenges mentioned in Section 3.2. For fairest comparison of the decomposition structure learning techniques, we modify DSCRDM to use its method of structure learning  on dictionary learned using PED and refer to the resulting method as \textit{Modified DSCRDM (MDSCRDM)}. \\
\textbf{Metrics:} We assess the performance of the different methods using three different metrics:  macro-averaged $f$ score ($Mf$), micro-averaged $f$ score ($\mu f$) and Normalized Disaggregation Error (NDE) which were introduced in ~\cite{kolter2012approximate}.  \par
We use True Positive (TP) as the portion of the predicted energy consumption of a device that it actually consumes, False Negative (FN) as the portion of the actual energy consumption that is not predicted and False Positive (FP) as the portion of predicted energy consumption that is not a part of the actual energy consumption. Since there is no notion of predicting non-consumption of energy, True Negative (TN) is 0.

\begin{equation}
  \text{TP}_{i,t} =  \frac{\min(x_{i,t}, \hat{x}_{i,t})}{ \hat{x}_{i,t}}, \text{   FP}_{i,t} = \frac{\min(0,  \hat{x}_{i,t} - x_{i,t})}{\hat{x}_{i,t}},
  \end{equation}
  \begin{equation}
  \text{FN}_{i,t} = \frac{\min(0,  x_{i,t}-\hat{x}_{i,t})}{x_{i,t}}, \text{TN} =0.
\end{equation}
For $\mu f$ score, we define the precision and recall as: 
\begin{equation}
 \mu Precision =\frac{\sum_{t=1}^T \sum_{i \in \mathcal{D}} \text{TP}_{i,t}}{\sum_{t=1}^T \sum_{i \in \mathcal{D}} (\text{TP}_{i,t}+ \text{FP}_{i,t}) } 
\end{equation}
\begin{equation}
 \mu Recall =\frac{\sum_{t=1}^T \sum_{i \in \mathcal{D}} \text{TP}_{i,t}}{\sum_{t=1}^T \sum_{i \in \mathcal{D}} (\text{TP}_{i,t}+ \text{FN}_{i,t}) } 
\end{equation}

For $M f$ score, we define the precision and recall as: 
\begin{equation}
    Precision_i = \frac{\sum_{t=1}^T \text{TP}_{i,t}}{\sum_{t=1}^T (\text{TP}_{i,t}+\text{FP}_{i,t})} ,  
\end{equation}
\begin{equation}
    Recall_i = \frac{\sum_{t=1}^T \text{TP}_{i,t}}{\sum_{t=1}^T (\text{TP}_{i,t}+\text{FN}_{i,t})} 
\end{equation}
\begin{equation}
   MPrecision = \frac{\sum_{i\in \mathcal{D}}Precision_i}{|\mathcal{D}|}, 
\end{equation}
\begin{equation}
   MRecall = \frac{\sum_{i\in \mathcal{D}}Recall_i}{|\mathcal{D}|}, 
\end{equation}
where, $\mathcal{D}$ is the set of devices in the house. The harmonic mean of $\mu$ and $M$ precision and recall  gives $\mu$ and $M$ $f$ scores ~\cite{tabatabaei2017toward}, respectively. 

The difference between the two $f$ scores is that the $M f$ score gives equal weight to each device, whereas the $\mu f$ score tends to give higher importance to devices that consume high power.\par

We also use the popular metric, Normalized Disaggregation Error\\ (NDE)~\cite{zhong2014signal}. 
\begin{equation}
NDE = \frac{\sum_{i,t} (\hat{x}_{i,t}-x_{i,t})^2}{\sum x^2_{i,t}}
\end{equation}
\textbf{Model Training and Parameter Selection:} We train each model using $80\%$ of the data with respect to time and test it on the rest. The parameters of the methods used are kept unchanged. 
We compute the performance of GDDM in terms of the $f$ scores and NDE averaged over all the houses of ECO and UK-Dale at different values $w\in\{5,15,25,35\}$  and $k\in\{20,40,60\}$.  We find that $w=15$ gives the best performance for the method and for $w=15$, the best performance on the test data is obtained when $k$ is $40$. For our device decomposition methods we use $w=15$ and $k=40$.
\par

For DPDDM, the weights associated with a node is calculated as $|\mathcal{S}|^\alpha$, where ${\mathcal{S}}$ is the set of devices belonging to the node. 
We try three different values of $\alpha [1,2,3]$ and find that $\alpha = 2$ gives the best performance for all the metrics.

\section{Results and Discussion}

\begin{table}[]
\centering
 \caption{ The performance of GDDM\_ADMM, GDDM\_IP the houses of the UK-Dale dataset.}
 \begin{tabular}{c c c c}
    \toprule
Method & $\mu f$ & $Mf$ & NDE    \\
\hline
GDDM\_IP & 0.517&0.761 & \textbf{1.665}\\
GDDM\_ADMM & \textbf{0.535}& \textbf{0.765} & 1.701 \\
\bottomrule
\end{tabular}
\end{table}
\begin{table*}[]\label{tab:uk_dale_performance}
\centering
\label{my-label}
\begin{threeparttable}
\caption{The performance of device decomposition methods for all the house in the UK-Dale and ECO dataset. Both our methods DPDDM, GDDM (equi-sized) outperform the existing state-of-the-art methods.  \tnote{1,2,3}}
\begin{tabular}{lccccccc}
\toprule
&\multicolumn{3}{c}{UK-Dale}  & & \multicolumn{3}{c}{ECO}  \\ \cmidrule{2-4} \cmidrule{6-8}
                                                                   
&  $\mu f$ & $M f$ & NDE & & $\mu f$ & $M f$ & NDE\\     
\midrule
GDDM (equi-sized) &\textbf{0.535}$\pm$\textbf{0.168} &\textbf{0.765}$\pm$\textbf{0.038}&\textbf{1.701}$\pm$\textbf{0.884}&&\textbf{0.473$\pm$0.139}&\textbf{0.741}$\pm$\textbf{0.070}& \textbf{1.018}$\pm$\textbf{0.617}\\ 
          
PED&0.409$\pm$0.209&0.696$\pm$0.053&2.832$\pm$1.131&&0.405$\pm$0.036&0.682$\pm$0.108&3.321$\pm$0.434 \\              
seq2point &0.432$\pm$0.300&0.681$\pm$0.108&1.753$\pm$1.445&&0.338$\pm$0.079&0.639$\pm$0.015&2.488$\pm$0.195 \\   
\bottomrule
\end{tabular}
\begin{tablenotes}

\item[1] shown are mean$\pm$ standard deviation values.
\item[2] \textbf{Bold} numbers are the best performance for the corresponding metric.
\item[3] For $M f$ and $\mu f$ measures higher values and for NDE lower values are better, respectively.

\end{tablenotes}
\end{threeparttable}

\end{table*}

\subsection{Comparison of optimization methods for decomposing the energy signal}
PED uses standard integer programming solver MOSEK~\cite{cvx}  for solving optimization functions with integer constraints as is the case in Equation (8). In order to validate that the improvement in the performance of the device decomposition methods is because of utilizing decomposition of device set scheme and not the optimization solver, we compare the results of GDDM with equi-sized partition with the integer programming solver (GDDM\_IP)  and GDDM with ADMM (GDDM\_ADMM). The results are shown in Table 2 and none of the methods perform consistently better than the other. The discussion on solvers for integer convex quadratic problem also suggests that SDP formulation achieves a good optimal solution~\cite{park2017semidefinite} comparable to standard integer programming solvers. 

\subsection{Comparison of device decomposition methods}

Table 3 shows the performance of GDDM (1-vs-rest), GDDM (equi-sized partition), DPDDM, and MDSCRDM for dividing the devices at each node on the ECO and the UK-Dale datasets. From the results, we observe that the equi-sized partitioning methods (GDDM (with equi-sized) and DPDDM) perform better than the 1-vs-rest (MDSRDM and GDDM (1-vs-rest)) partitioning methods. We perform a paired t-test on the values of $f$ scores and NDE obtained for all the houses in both the datasets. For the null hypothesis that the results of GDDM (equi-sized partition) and MDSCRDM are derived from the same distribution, the resulting p-value (~ 0.01) shows that the difference is statistically significant. The same holds for results of GDDM (equi-sized partition) and GDDM (1-vs-rest partition).

\par

 
\begin{table*}[]
\centering
\begin{threeparttable}
\caption{Per appliance performance comparison. This table shows how our method performs at predicting the power consumption of each appliance. Our method performs better for the appliances which consume most power in a household and are used frequently (Fridge and freezer). }
\begin{tabular}{lccccccccccc}
\toprule
\multicolumn{1}{l}{} & \multicolumn{3}{c}{$\mu f$}   &                                                   & \multicolumn{3}{c}{$M f$}  &                                                     & \multicolumn{3}{c}{NDE}                                                            \\ \cmidrule{2-4} \cmidrule{6-8} \cmidrule{10-12}
Appliance             & \multicolumn{1}{l}{GDDM} & \multicolumn{1}{l}{PED} & \multicolumn{1}{l}{seq2point} & &\multicolumn{1}{l}{GDDM} & \multicolumn{1}{l}{PED} & \multicolumn{1}{l}{seq2point} & & \multicolumn{1}{l}{GDDM} & \multicolumn{1}{l}{PED} & \multicolumn{1}{l}{seq2point} \\
\hline
Fridge                & \textbf{0.620}            & 0.181                   & 0.445                     &    & \textbf{0.763}            & 0.235                   & 0.704            &             & \textbf{0.552}            & 0.894                   & 0.997                         \\
Dryer                 & 0.426                     & 0.149                   & \textbf{0.447}     &           & \textbf{0.653}            & 0.246                   & 0.607      &                   & \textbf{1.005}            & 1.029                   & 1.093                         \\
Coffee Machine        & \textbf{0.037}            & 0.001                   & 0.027                      &   & 0.835                     & \textbf{0.945}          & 0.085                &         & 1.896                     & 1.354                   & \textbf{1.003}                \\
Kettle                & \textbf{0.009}            & 0.000                   & 0.003                   &      & 0.946                     & \textbf{0.991}          & 0.826                  &       & 1.976                     & 2.946                   & \textbf{1.037}                \\
Washing Machine       & 0.001                     & 0.004                   & \textbf{0.005}      &          & 0.844                     & \textbf{0.941}          & 0.738               &          & \textbf{0.999}            & 1.644                   & 1.237                         \\
PC                    & 0.047                     & 0.013                   & \textbf{0.068}    &            & 0.441                     & \textbf{0.810}          & 0.619         &                & 1.655                     & \textbf{1.026}          & 1.046                         \\
Freezer               & \textbf{0.542}            & 0.174                   & 0.273                    &     & \textbf{0.714}            & 0.423                   & 0.593              &           & \textbf{0.926}            & 0.980                   & 5.879                         \\
\hline
Average               & \textbf{0.240}            & 0.075                   & 0.181                      &   & \textbf{0.742}            & 0.656                   & 0.596              &           & \textbf{1.287}            & 1.410                   & 1.756   \\
\bottomrule
\end{tabular}
\begin{tablenotes}
\item[1] \textbf{Bold} numbers are the best performance for the corresponding metric.
\item[2] For $M f$ and $\mu f$ measures higher values and for NDE lower values are better, respectively.
\end{tablenotes}
\end{threeparttable}
\end{table*}
\subsection{Comparison with PED and Seq2point}

Tables 3 shows the performance achieved by GDDM, Seq2point, and PED on the ECO and UK-Dale datasets. For ECO, GDDM improves average performance by $16.7\%$, $8.65\%$ and $69.3\%$ in terms of the $\mu f$, $M f$ and NDE, respectively compared to PED. Compared to the seq2point method, we obtain an improvement of $39.9\%$, $15.9\%$ and $59.3\%$ in terms of the  $\mu f$, $M f$ and NDE, respectively. The improvements obtained by GDDM is statistically significant at confidence level of $1\%$. In comparison with  PED, for UK-Dale, GDDM improved the performance by $30.8\%$, $10\%$ and $39.9\%$, in terms of $\mu f$, $M f$ and NDE, respectively. Compared to seq2point method, GDDM achieved performance improvement of $23.8\%$, $12.3\%$, and $2.9\%$ in terms of $\mu f$, $M f$ and NDE, respectively on UK-Dale. The paired t-test shows that the improvement in terms of $M f$ and $\mu f$ is significant (pvalue $<0.02$) while the improvement on  NDE is not.   
Additionally, we compare the three methods on appliance level performance. Table 4 shows the performance of various methods on all the appliances of house 1 of the ECO dataset.  We can see that our method GDDM performs better than the other methods for half the number devices in terms of $\mu f$. The $M f$ score of PED method is better for $4$ devices compared to our method. In terms of NDE, our method performs better for $4$ devices compared to the other two methods. Additionally, our method shows better performance in terms of the overall average $\mu f$, $M f$  and NDE than the other two. 

\subsection{Visualization of Device Decomposition Strucuture}
\begin{figure*}
\centering
\begin{subfigure}[b]{0.5\linewidth}
\includegraphics[width=\linewidth]{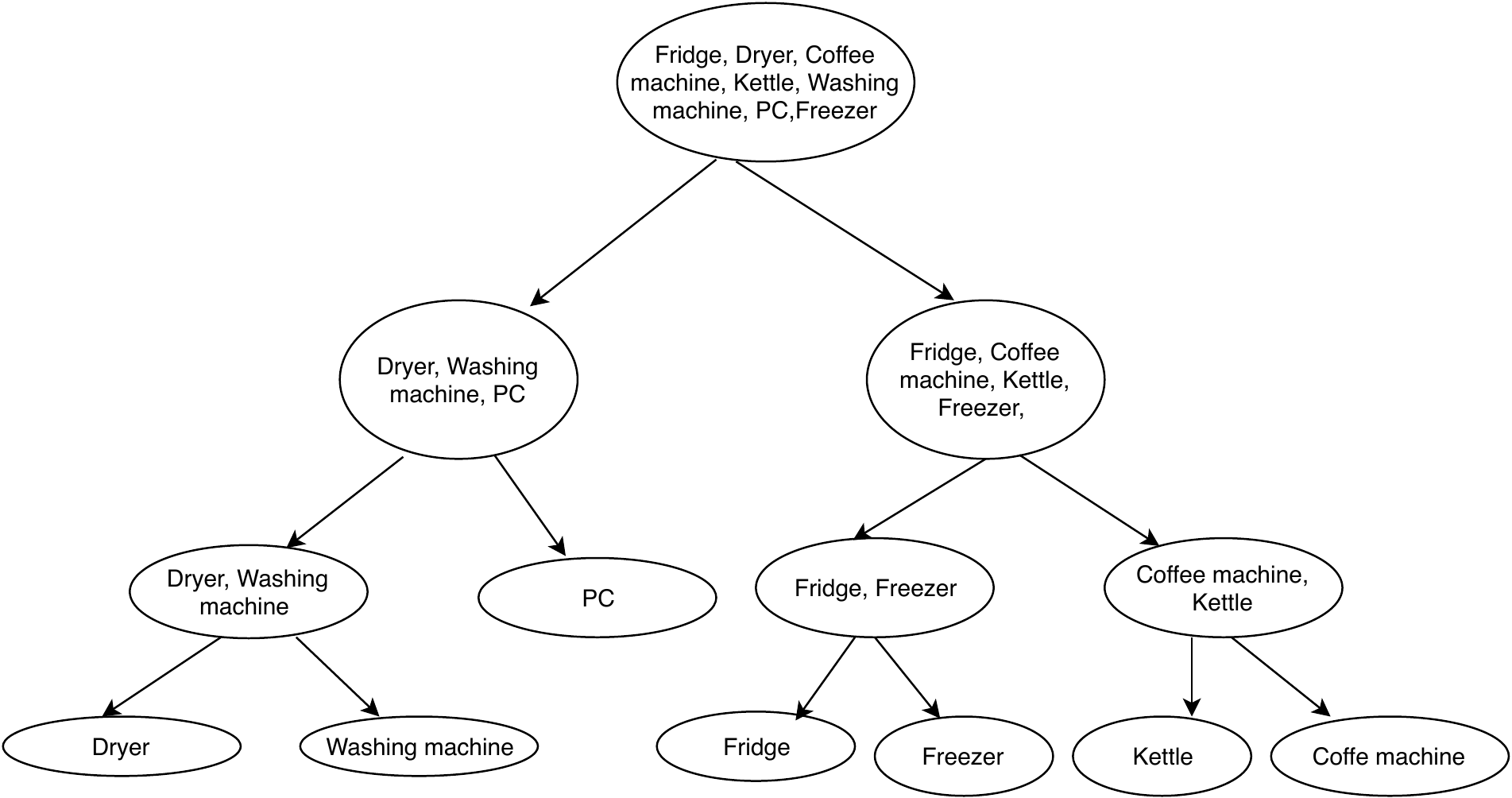}
\subcaption{Decomposition structure used for disaggregating the electricity consumption of House 1 in ECO dataset. The devices such as (Kettle, Coffee machine), which co-occur are discovered by our method and used for disambiguating similar power consuming devices.}\label{tree}
\end{subfigure}
\begin{subfigure}[b]{0.6\linewidth}
\includegraphics[width=\linewidth]{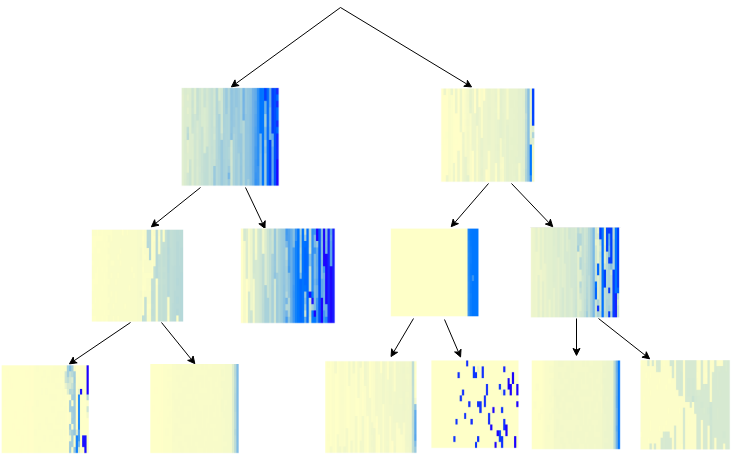}
\subcaption{Visualization of powerlets of the pseudo devices at each level of the device decomposition structure shown above. }\label{heatmap}
\end{subfigure}
\caption{Visualization of device decomposition structures. }
\label{decompose}
\end{figure*}

In order to show the effectiveness of designing a decomposition structure for performing disaggregation, we visualize how disaggregation is performed on the $1$st household of the ECO dataset. Figure~\ref{tree} shows the device decomposition structure used by our method for performing disaggregation. Note that even though appliances such as washing machine and fridge show similar power consumption values when active, they can be distinguished by our method as the powerlets of pseudo devices \{PC, Dryer and washing machine\} or \{Fridge, Freezer, Kettle and coffee machine \} are different. This is because in some cases the activation of fridge could co-occur with freezer or kettle. Thus, using decomposition structure helps in capturing the co-occurence of devices and improve the disaggregation performance for this house. Figure~\ref{heatmap} visualizes the powerlets of different (pseudo) devices at each level.\par

\begin{table}[]
\centering
\label{time}
 \caption{ Training time (in minutes) of various methods for training \text{house2} of UK Dale dataset.}
 \begin{tabular}{r r r r r}
    \toprule
GDDM & DPDDM &MDSCRDM & PED & seq2point     \\
\hline
3.33&8.56 & 323.33 & 0.33 & 1735.45\\

\bottomrule
\end{tabular}
\end{table}
\subsection{Training Efficieny}
Table 5 shows the training time taken by various methods using Intel Xeon E7 processor. PED takes the least time for training while seq2point takes the most. Our approaches take an order of magnitude less time to train as compared to MDSCRDM because MDSCRDM solves an optimization function at each level for splitting the device set while our methods use heuristics to find the best split. 
\section{Conclusion and Future work}

In this paper, we improved the performance of the existing methods to obtain appliance level energy consumption from the aggregated energy of a house. Our methods leverage the fact that devices operate concurrently at specific operation modes. In order to capture the concurrent operating devices, we find the representatives of power consumption of sets of devices. We decompose the device set hierarchically to improve the performance of the task. We have empirically shown that device decomposition methods mostly outperform the state of the art methods on two real world data sets. The strength of our methods is that they are able to learn the devices that operate concurrently and use this knowledge to distinguish two similar energy consuming devices. \par

As part of future work, we plan to model the power consumption pattern of devices by including how the power consumption of device vary over time. Devices usually exhibit complex exponential decays or growth, bounded min-max, and cyclic patterns~\cite{iyengar2016non} of power consumption. This idea can be incorporated in extracting the powerlets. Additionally, a limitation of our model is that it is heavily dependent on the idea that  devices in a house co-occur at some of their operation modes. However, if that is not the case, we can use simpler methods, such as PED. In future we plan to design a method to detect if some devices in the set go together and use our method to disaggregate energy only for those scenarios, otherwise use PED.
\section{Acknowledgement}
This work was supported in part by NSF (1447788, 1704074, 1757916, 1834251), Army Research Office (W911NF1810344), Intel Corp, and the Digital Technology Center at the University of Minnesota. Access to research and computing facilities was provided by
the Digital Technology Center and the Minnesota Supercomputing Institute.
\bibliographystyle{ACM-Reference-Format}

\bibliography{main.bib}
\end{document}